# Machine Learning Enables Ultra-Compact Integrated Photonics Through Silicon-Nanopattern Digital Metamaterials

Sourangsu Banerji[1], Apratim Majumder[1], Alex Hamrick[2], Rajesh Menon[1], and Berardi Sensale-Rodriguez[1, *]

[1] Department of Electrical and Computer Engineering, University of Utah, Salt Lake City, UT 84112, USA

[2] School of Computing, University of Utah, Salt Lake City, UT 84112, USA

* E-mail: berardi.sensale@utah.edu

**Abstract**

In this work, we demonstrate three ultra-compact integrated-photonics devices, which are designed via a machine-learning algorithm coupled with finite-difference time-domain (FDTD) modeling. Through digitizing the design domain into "binary pixels," these digital metamaterials are readily manufacturable as well. By showing a variety of devices (beamsplitters and waveguide bends), we showcase the generality of our approach. With an area footprint smaller than $\lambda_0^2$, our designs are amongst the smallest reported to-date. Our method combines machine learning with digital metamaterials to enable ultra-compact, manufacturable devices, which could power a new "Photonics Moore's Law."

## 1. Introduction

Silicon photonics has witnessed major breakthroughs in recent years [1]. A critical enabler for its unprecedented success can be attributed to device designs that demonstrated favorable characteristics like high sensitivity, low-loss, and high index-contrast [1-5]. CMOS-compatible integrated-photonic systems have also contributed to this progress [1, 2], and have led to the miniaturization of optical circuits. In contrast to integrated electronics, scalable design methods for ultra-dense integrated photonics are still missing [6, 7].

Conventional nanophotonic design is often based upon theoretical intuition [8-11]. This approach has two drawbacks. First, its application is constrained to known structures and well modeled light-matter interactions [12]. Second, device designs based on such methods might not satisfy practical performance requirements like compactness, efficiency, bandwidth, and low-loss. In order to solve these problems, a variety of numerical approaches have been developed. These include inverse-design [13], evolutionary algorithms [14], objective-first algorithms [15-18], topology optimization [19], and direct-binary-search algorithms [20-26]. The latter approach utilized digital metamaterials (DMMs), where the design domain is discretized in sub-units (pixels) based upon the fabrication capabilities, which allows for robust, manufacturable devices. Recently, machine learning (ML) has been applied to this design problem [27-31]. ML has proven to be a powerful design approach primarily due to (a) easy hardware parallelization, (b) relative independence from the choice of initial solutions, and (c) potential for generating manufacturable designs.

Recently, we applied an ML method, binary-Additive Reinforcement Learning Algorithm (b-ARLA), to design one of the smallest T-junction splitters reported to date [5]. In order to showcase the general nature of this design approach, here we extend b-ARLA to design three fundamental

DMM devices: a 50:50 Y-junction beamsplitter (**Fig. 1a**), and 90° (**Fig. 1b**) and 180° (**Fig. 1c**) waveguide bends.

The other contributions of this work include a careful analysis of the impact on device performance of (*i*) the number of guesses during training, that is the trade-off between performance and computational cost; and (*ii*) the dimension of the digitized pixels, which define the number of degrees of freedom. We acknowledge that the devices reported here show among the smallest footprints, but not necessarily the best performance (for instance, the insertion loss is higher than that of other devices proposed in the literature). However, these devices are presented as examples of our design approach, and further improvements in the algorithm and loss function formulations could mitigate these deficiencies.

## 2. Design and Optimization

Our algorithm (b-ARLA) combines both the "*additive updates*" feature of a perceptron algorithm as well as the "*reward for state idea*" associated with reinforcement learning [**5, 33**]. The flow diagram of the algorithm is shown in **Fig. 2**. The algorithm essentially consists of two stages: training (highlighted in red) and inference (highlighted in green). Further details about the algorithm and its implementation are provided in our previous work [**5**].

The training stage initiates with the creation of a set of random designs (guesses). In the examples reported here, we considered square domains (1.2 μm × 1.2 μm in size) comprised of N × N pixels. Each pixel is randomly assigned to be either air (refractive index =1) or Silicon (refractive index = 3.46). The input and output waveguides are: 1 μm (length) × 0.44 μm (width) × 0.22 μm (height). The thickness of the substrate is 2 μm. Therefore, for N = 12, the size of each pixel is 0.1 μm × 0.1 μm. A 2.5D varFDTD (variational FDTD, Lumerical) method [**34**] calculated the steady-state electromagnetic fields at a wavelength, $\lambda_0$ = 1.55 μm. TE polarization (with non-zero components of $E_x$, $E_y$, and $H_z$) was assumed. Perfectly matched layers (PML) surrounded the boundaries of the computational domain. During the training as well as the inference stages, Lumerical MODE solutions was employed to extract the steady-state response of the structures with 2.5D varFDTD. An additional post-validation after the inference stage was done to crosscheck the obtained results across a full 3D FDTD with Lumerical FDTD solutions. The insertion loss, defined as the ratio (in dB) of power transmitted through the device to the incident power was computed, which is our metric for optimization.

First, we analyzed the predicted insertion loss for the Y-junction beamsplitter as a function of the number of guesses employed during training (**Fig. 3a**). The final prediction of insertion loss decreases as the number of random designs employed in the training stage is increased. Moreover, from the slope of the curve is seen that the improvement in the solution gradually drops as the number of guesses is increased. That is, from a certain point, around computational cost would not significantly pay off in benefits. From this perspective, we chose 10,000 guesses as a reasonable trade-off between computational cost and predicted device performance.

Next, we analyzed the impact of the number of pixels on the predicted insertion loss for the Y-junction beamsplitter. For this purpose, we ran the algorithm discretizing the domain geometry in 4 × 4 (N = 4), 6 × 6 (N = 6), and 12 × 12 (N = 12) pixels. As expected, the predicted insertion loss decreases with the number of pixels (**Fig. 3b**).

## 3. Results and Discussion

**Figure 3(c-e)** shows the insertion loss during design for the three devices. The value for each guess is shown as red dots, while that of the final device is in green (2D var FDTD) and blue (3D

FDTD). The achieved insertion losses are 0.92dB, 1.07 dB, and 1.73 dB for the beamsplitter, 90˚ bend, and 180˚ bend, respectively. The symmetry of the beamsplitter likely allows for fewer guesses and better performance (**see Fig. 2**).

The steady-state-field distribution of the Y-junction beamsplitter is shown in **Fig. 4a**. Although the device was designed for one wavelength, its operating bandwidth is quite broad (1.54 μm to 1.56 μm), as depicted in **Fig. 4b**. The field-distribution confirms efficient splitting of the input mode and strong modal match at the output waveguide, resulting in good transmission efficiency. **Figures 4c** and **4e** show the steady-state-field distributions for the 90˚ and 180˚ bends, respectively. Similar operating bandwidths, as in the beamsplitter, are also observed (see **Figs. 4d** and **4f**).

We note that larger bend devices (2 orders of magnitude larger than our devices) with somewhat better performance have been reported [**35**]. Furthermore, the bend devices reported here are 60% smaller than the smallest bend device reported before [**36**] and e.g., for the 180º bend provide for a remarkably small bend radius of just ~$\lambda_o$/10 (the input and output waveguides are spaced by 0.32 μm, which corresponds to an effective bend radius of 0.16 μm ~ $\lambda_o$/10).

## 4. Conclusion

We proposed a general design framework for ultra-compact digital metamaterial devices through the employment of a machine-learning algorithm, b-ARLA. This approach is general and can be applied to both passive and active devices. We designed the of the smallest in-class devices: beamsplitters and waveguide bends reported so far with a footprint of only 1.2 μm × 1.2 μm thus < $\lambda_0^2$. The possibility of designing and integrating such ultra-compact and efficient nanophotonic structures will allow to increase the density of on-chip photonic components dramatically and

therefore enable more complex photonic integrated circuits, thereby enabling a new "Photonics Moore's Law."

**Acknowledgments**

This work was supported by the NSF awards: ECCS # 1936729 and MRI #1828480.

**Figures**

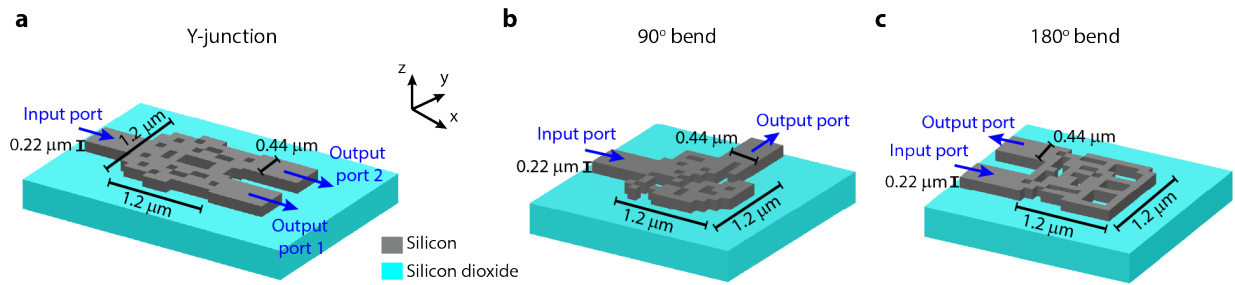

**Fig. 1.** Using machine learning to design digital metamaterials devices: **(a)** 50:50 beamsplitter (Y-junction). Light from the input port is split equally between the two output ports. **(b)** 90°, and **(c)** 180° waveguide bends, where light from the input port is routed to the output port. All three devices have a size of 1.2 μm × 1.2 μm, and the design wavelength is λ = 1.55 μm.

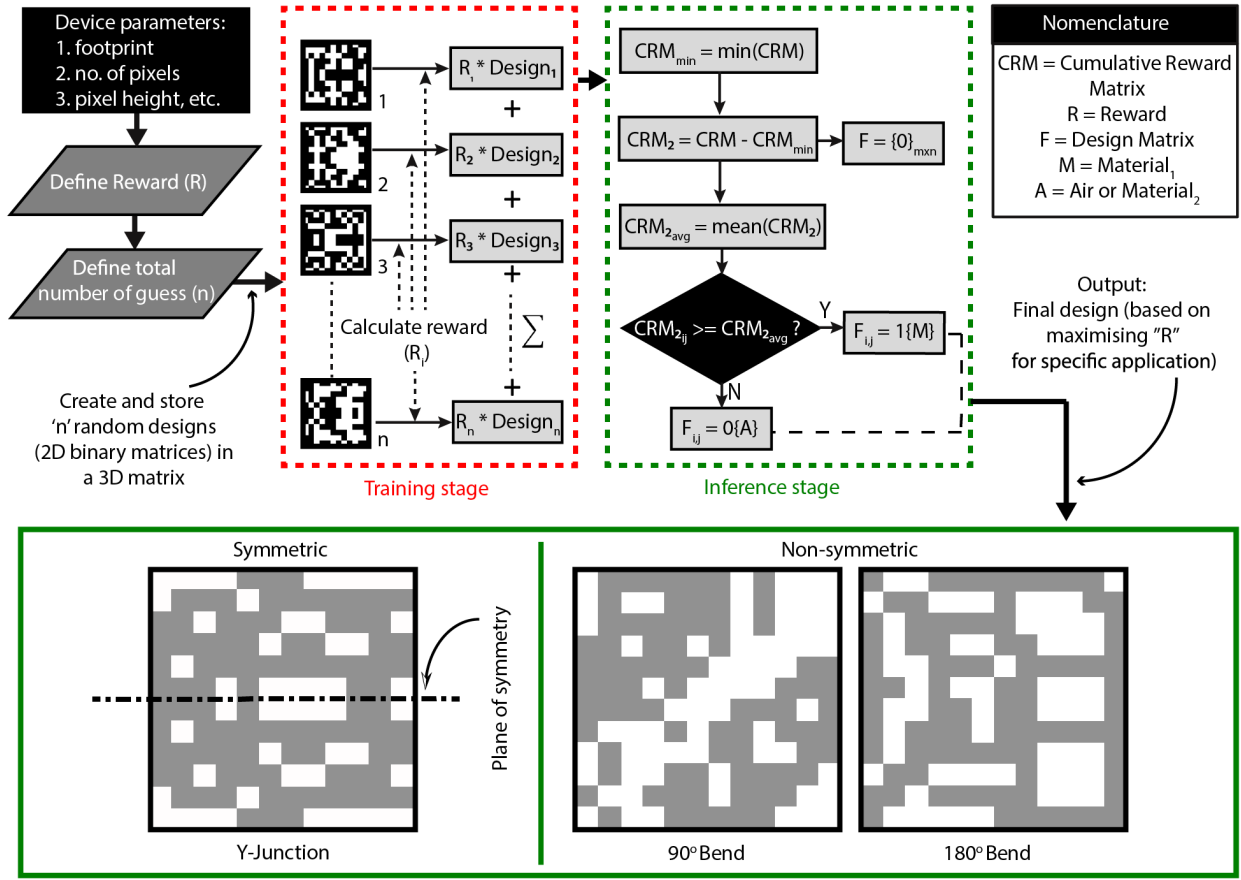

**Fig. 2.** Flow diagram of the binary-Additive Reinforcement Learning Algorithm (b-ARLA). The algorithm bases its output, i.e., the final design that maximizes the given reward defined during the pre-training stage. This is advantageous since the inference is one-time and does not depend on whether the resultant structure is symmetric or not. (Figure adapted from [5])

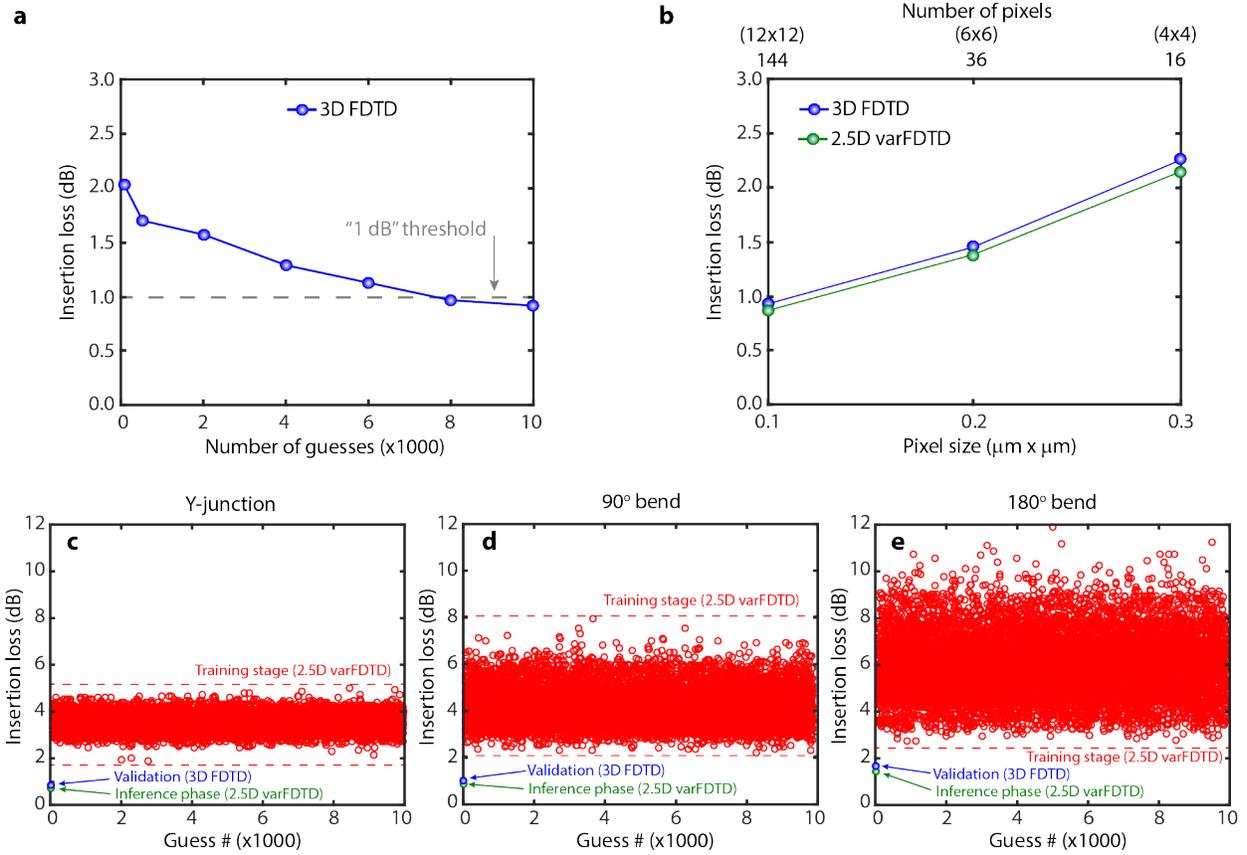

**Fig. 3. (a)** The final prediction of the insertion loss (in dB) as a function of the total number of guesses used in the inference stage for the Y-junction. **(b)** Insertion loss achieved in the final prediction for the Y-junction with 0.1, 0.2, and 0.3 μm pixel sizes. The smallest the pixel size (that is, the more number of pixels, e.g., degrees of freedom), the better the final performance. **(c-e)** The insertion loss (in dB) for each guess (hollow blue dots) and final prediction (solid red dot) using 10,000 guesses across all the designed nanophotonic devices. The smallest insertion loss achieved are **(c)** ~0.92 dB (beamsplitter), **(d)** ~1.07 dB (90° bend) and **(e)** ~1.73 dB (180° bend).

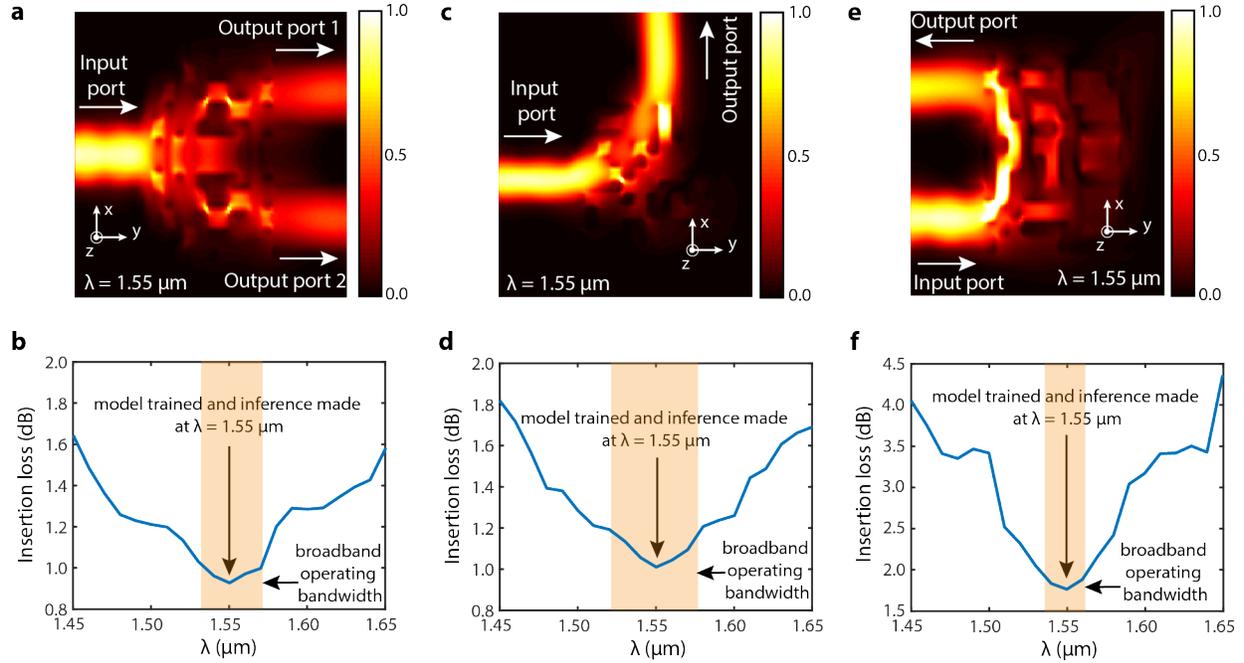

**Fig. 4.** The steady-state response at the operational wavelength of 1.55 μm for the (a) Y-junction splitter, **(c)** 90° and **(e)** 180° waveguide bend; and the insertion loss for the structures under broadband operation (1.50 - 1.60 μm) for each design: **(b)** Y-junction splitter, **(d)** 90°, and **(f)** 180° waveguide bends.